\documentclass[aps,pra,showpacs,epsfigure,twocolumn]{revtex4}
\usepackage{dcolumn}    
\usepackage{bm} 
\usepackage{graphicx}
\usepackage{amsmath}    
\usepackage{latexsym}
\usepackage{amsfonts}   
\usepackage{amssymb}
\usepackage{array}      
\usepackage{epsfig}
\usepackage{times}

\newcommand{\ket}[1]{\left\vert#1\right\rangle}

\newcommand{\bra}[1]{\left\langle#1\right\vert}

\begin{document}

\title{Non-classicality of optomechanical devices in experimentally realistic operating regimes.}
\date{\today}
\author{G. Vacanti$^1$, M. Paternostro$^2$, G. M. Palma$^3$, M. S. Kim$^4$, and V. Vedral$^{1,5,6}$ }
\address{$^1$Centre for Quantum Technologies, National University of Singapore, 3 Science Drive 2, Singapore\\
$^2$Centre for Theoretical Atomic, Molecular, and Optical Physics, School of Mathematics and Physics, Queen's University, Belfast BT7 1NN, United Kingdom\\
$^3$NEST Istituto Nanoscienze-CNR and Dipartimento di Fisica, Universit\'{a} degli Studi di Palermo, via Archirafi 36, I-90123 Palermo, Italy\\
$^4$QOLS, Blackett Laboratory, Imperial College London, London SW7 2BW, United Kingdom\\
$^5$Clarendon Laboratory, University of Oxford, Parks Road, Oxford OX1 3PU, United Kingdom\\
$^6$Department of Physics, National University of Singapore, 2 Science Drive 3, Singapore}
\begin{abstract}
Enforcing a non-classical behavior in mesoscopic systems is important for the study of the boundaries between quantum and classical world. Recent experiments have shown that optomechanical devices are promising candidates to pursue such investigations. Here we consider two different setups where the indirect coupling between a three-level atom and the movable mirrors of a cavity is achieved. The resulting dynamics is able to conditionally prepare a non-classical state of the mirrors  by means of projective measurements operated over a pure state of the atomic system. The non-classical features are persistent against incoherent thermal preparation of the mechanical systems and their dissipative dynamics.
\end{abstract}

\maketitle

\section{Introduction}

Interesting experimental endeavors have recently challenged the widely-accepted assumption that quantumness is an exclusive prerogative of microscopic and isolated systems. These efforts show that complex and large objects comprising many elementary constituents or endowed with a variety of degrees of freedom can display important non-classical features~\cite{naik,thompson,schliesser1,corbitt,schliesser2,schliesser3,oconnell,demartini,sekatski}. In general, quantum control under unfavorable operating conditions is an important milestone in the study of the quantum-to-classical transition and, as such, should be pursued to achieve a better understanding of the conditions enforcing and implying quantum mechanical features in the state of a given system.  This topic has recently  become  the focus of an intense research activity, at all levels, boosted by the ability to experimentally manipulate systems composed of subparts having variegated nature. We can now coherently control the interaction between radiation and Bose-Einstein condensates~\cite{Brennecke,colombe} while mesoscopic superconducting devices compete with atoms and ions for the realization of cavity quantum electrodynamics~\cite{schoelkopf,majer,Wallraff}. Equally remarkable is the progressive entering of purely mechanical systems into the realm of experimental controllability~\cite{gigan,arcizet1,kleckner,arcizet2,Groblacher1,Groblacher2,chan}. The operative conditions and the intrinsic nature of the systems involved in these examples often deviate from the naive requirements for ``quantumness'': ultra-low temperatures, full addressability and ideal preparation of the system. The design and exploitation of such interesting setups is giving further emphasis to investigations performed along the lines of the question raised above~\cite{marshall,ferreira,mauro}. 

Here, we prove how  non-classical behaviors can be induced in  massive mesoscopic  systems out of the reach of direct addressability.  The indirect interaction with a fully controllable microscopic system enforces  non-classical mesoscopic states, robust against adverse operative conditions (such as temperature). Our study is performed in the micro-scale domain and involves two different  optomechanical cavity-quantum electrodynamics settings. It proposes a scenario for the observation of induced non-classical features, such as non-local correlations  and negative values of the Wigner function,  that are truly mesoscopic (thus different from more extensively studied nano-scale setups~\cite{armour,rabl,tian,rodrigues}), well-controllable and, although close to experimental capabilities in the fields of optomechanics and light-matter interaction, yet unexplored.

The paper is organized as follow: in Sec.~\ref{SM}, we discuss a setup in which one mesoscopic object (a movable end-mirror of an optical cavity) interacts with a microscopic system (a three-level atom) through the radiation inside the cavity. In this context, we  study the correlations between the two systems as well as the non-classical features induced on the state of the mirror. In Sec.~\ref{TM} we extend our analysis to a system where both cavity  mirrors  interact with  the atom. This setup allows us to investigate the correlations between two truly mesoscopic systems, revealing how quantum effects can survive to adverse environmental conditions such as dissipation and thermalization.
\begin{figure}[b]
\includegraphics[width=0.42\textwidth]{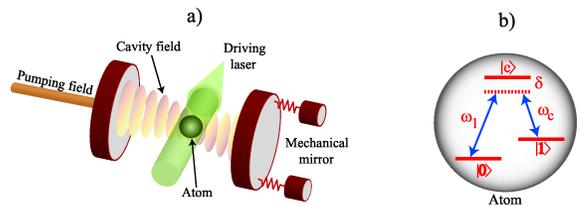}
\caption{{ (Color online) \bf (a)} Scheme of the system. {\bf (b)} Energy levels of the atom driven by an off-resonant two-photon Raman transition.} \label{scheme}
\end{figure} 

\section{Single Mirror}\label{SM}
Here we consider an optomechanical system consisting of a cavity whose end-mirror can oscillate under the action of the radiation-pressure force. A three-level atom is placed inside the cavity and the system parameters are chosen so that an effective atom-mirror coupling is achieved. We show how the state of the system  reveals strong non-classical features such as non-local correlations between the atom and the mirror and negative values of the Wigner function of the mirror, even in presence of dissipative processes and non-zero temperature.

\subsection{The Model}
\label{subsectionOMModel}
The system that we consider involves a three-level atom in a $\Lambda$ configuration, coupled to a single-mode optical cavity pumped by an laser field at frequency $\omega_p$ and with a movable mirror. The atom is driven by a second external field at frequency $\omega_i$ that enters the cavity radially (see Fig.~\ref{scheme}). We label $\{\ket{0},\ket{1}\}$ the states belonging to the fundamental atomic doublet and $\ket{e}$ the excited state. The atomic transition $|0\rangle{\leftrightarrow}|e\rangle$  is guided, at rate $\Omega,$  by the external  field at frequency $\omega_i$. On the other hand, the transition $|1\rangle{\leftrightarrow}|e\rangle$ is coupled to the cavity field at frequency  $\omega_c$ with coupling constant $g$. We call $\delta$ the detuning between each transition and the respective driving field, while $\Delta{=}\omega_c{-}\omega_p$ is the cavity-pump detuning. The movable mirror is modeled as a harmonic oscillator with frequency $\omega_m$, coupled to the cavity field through radiation-pressure. We assume large single-photon Raman detuning and negligible decay rate $\gamma_e$ from the atomic excited state, so that $\delta\gg{\Omega,g}\gg\gamma_e$ and an off-resonant two-photon Raman transition is realized. Moving to an interaction picture defined by the operator $\omega_p\hat{a}^\dag\hat{a}+\omega_i\ket{e}_a\!\bra{e}+\omega_{10}\ket{1}_{a}\!\bra{1},$ the Hamiltonian of the overall system reads [we set $\hbar{=}1$ throughout the paper] $\hat{\cal H}_{\rm sys}=\hat{\cal H}_{a}+\hat{\cal H}_{R}+\hat{\cal H}_{m}+\hat{\cal H}_{c}+\hat{\cal H}_{mc}+\hat{\cal H}_{cp}\label{HTot}$, where 
\begin{equation}
\begin{split}\label{HTot}
&\hat{\cal H}_{a}\!=\!{\delta}\ket{e}_a\!\bra{e},\hat{\cal H}_{m}=\omega_m\hat{b}^\dag\hat{b},\\
&\hat{\cal H}_{c}=-\Delta\hat{a}^\dag\hat{a}, \hat{\cal H}_{mc}=\chi\hat{a}^\dag\hat{a}(\hat{b}+\hat{b}^\dag),\\
&\hat{\cal H}_{R}\!=\! \Omega(\ket{e}_a\!\bra{0}+\ket{0}_a\!\bra{e})+{g}(e^{i\Delta{t}}\hat{a}^\dag\ket{1}_a\!\bra{e}+h.c.)\\
\end{split}
\end{equation}
Here, $\hat{\cal H}_{a}$ is the atomic energy, $\hat{\cal H}_{R}$ is the Raman coupling, $\hat{\cal H}_m$ ($\hat{\cal H}_c$) is the mirror (cavity) free Hamiltonian and  $\hat{\cal H}_{mc}$ is the radiation-pressure term \cite{law} (with coupling rate $\chi$), where $\hat{a}$ ($\hat{a}^\dag$) is the annihilation (creation) operator of the cavity field and $\hat{b}$ ($\hat{b}^\dag$) is the corresponding operator of the mirror. Finally, $\hat{\cal H}_{cp}$ is  the cavity-pump interaction~\cite{walls}.  The pumping field ensures that a few photons are always present in the cavity, allowing a mediated interaction between the atom and the mirror. On the other hand, the purpose of the external field with rate $\Omega$ is to trigger the passages between the excited level $|e\rangle$ and the ground level $|0\rangle.$

If we further assume $\Delta\gg{g,\chi}$, both the atomic excited state and the cavity field are virtually populated and they can be eliminated from the dynamics of the system. This leads to the effective interaction Hamiltonian 
\begin{equation}
\hat{\cal H}_{\rm eff} = \eta\ket{0}_a\!\bra{0}(\hat{b}^\dag+\hat{b})
\label{HEffOneMirr}
\end{equation}
where  $\eta={\chi{g}^2\Omega^2}/{\delta^2\Delta^2}.$ The form of the effective coupling rate $\eta$ shows that all the considered coupling mechanisms are necessary in order to achieve the atom-mirror coupling. Through the two-photon Raman transition, the virtual quanta resulting from the atom-cavity field interaction are transferred (by the bus embodied by the cavity field) to the mechanical system. As a consequence, the state of the latter experiences a displacement (in phase space) conditioned on the state of the effective two-level atomic system resulting from the elimination of the excited state. $\hat{\cal H}_{\rm eff}$ involves the position quadrature operator $\hat{q}\propto\hat{b}+\hat{b}^\dag$ of the movable mirror. It is worth noticing that, if the cavity is driven by a bichromatic pump with frequencies $\omega_{p}$ and $\omega_{p}+\omega_m$ and a relative phase $\phi$, the effective coupling between the atom and the movable mirror can be made {\it flexible} in the sense that $\hat{q}$ is replaced by $\hat{b}e^{i\phi}+\hat{b}^\dag{e}^{-i\phi}$, making possible the displacement in any direction of the phase space of the movable mirror~\cite{cam,noiGP,phase,Leibfried}.

\subsection{Atom-Mirror Entanglement}
\label{subsecOMEnt}
We now focus on the quantification of  microscopic-macroscopic correlations between the atom and the mirror. First, we assume that the initial state of the movable mirror is a coherent state $\ket{\alpha}_m$ with amplitude $\alpha\in\mathbb{C},$ while the atom is assumed intially in $\ket{+}_a=(\ket{0}+\ket{1})_{a}/\sqrt{2}$. Under the action of the effective  Hamiltonian in Eq. (\ref{HEffOneMirr}), the initial state evolves into $\ket{\psi(t)}=\hat{\cal U}_t\ket{+,\alpha}_{am}$, where
\begin{equation}
\ket{\psi(t)}=\frac{1}{\sqrt{2}}(\ket{1,\alpha}+e^{-i\varPhi(t)}\ket{0,\alpha-i\eta t e^{-i\phi}})_{am}
\label{catstate}
\end{equation}
with $\varPhi(t)=\eta t\text{Re}[\alpha{e}^{i\phi}]$ and $\hat{\cal U}_t\equiv{e}^{-i\hat{\cal H}_{\rm eff} t}=\ket{1}_a\!\bra{1}\otimes\openone+\ket{0}_a\!\bra{0}\otimes \hat{D}(-i\eta t e^{i\phi})$, where $\hat{D}(\zeta) = e^{\zeta \hat{b}^\dag- \zeta^*\hat{b}}$ is the single-mode displacement operator~\cite{walls}. Eq.~(\ref{catstate}) is, in general, an entangled state of a microscopic and a mesoscopic system: its Von Neumann entropy depends on the value of $\eta t$ only. Intuitively, the larger the phase-space distance between $\ket{\alpha}$ and $\ket{\alpha-i\eta t}$, the closer the evolved state to a balanced superposition of bipartite orthogonal states, thus maximizing the entanglement. To give a figure of merit, for $\eta t=0.82$  the entropy is~$\sim0.8,$ while for $\eta t>1.7$ the entropy is  $>0.996$. Interestingly, the kind of control over the mirror state reminds of the ``quantum switch'' protocol for microwave cavities~\cite{davidovicharoche}, although here it is achieved over a truly mesoscopic device.
\begin{figure}[t]
\includegraphics[width=0.42\textwidth]{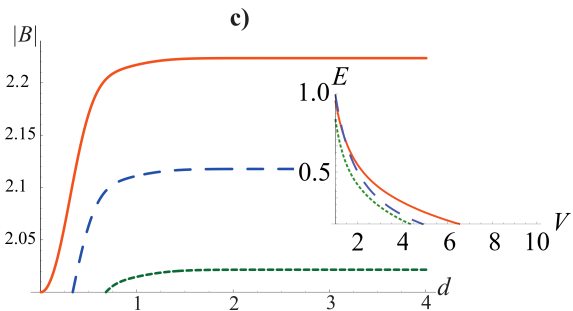}
\caption{(Color online) Maximum violation of the Bell-CHSH inequality against the displacement $d$. From top to bottom, the curves correspond to $V=1,3,5$ with $\eta t=2d$ and $\theta_1\simeq{3}\pi/2$ and are optimized with respect to $\theta$. The inset shows, from top to bottom, the logarithmic negativity $E$ against $V$ for projected states with $p=0,1$ and $2$, for $d=2$.} \label{CHSHoneMirr}
\end{figure} 

Although impressive progresses have recently been accomplished in active and passive cooling of micro- and nano-mechanical oscillators~\cite{chan}, it is realistic to expect the  mirror to be affected by thermal randomness due to its exposure to the driving field and/or to a phononic background at temperature $T$.  Exploiting the handiness of Eq.~(\ref{catstate}), we write the initial state of the mirror at thermal equilibrium (temperature $T$) and displaced by $d$ (due to the external pump) as 
\begin{equation}
\varrho_{m}^{\rm th}=\int{d}^2\alpha{\cal P}(\alpha,V)\ket{\alpha}_m\!\bra{\alpha}
\label{thermalstate}
\end{equation}
with
${\cal P}(\alpha,V)=\frac{{2}e^{-\frac{2|\alpha-d|^2}{V-1}}}{{\pi(V-1)}}$, $V=\coth (\omega_m/2k_bT)$ and $k_b$ the Boltzmann constant. Under $\hat{\cal U}_t$, the state $\ket{+}_a\!\bra{+}\otimes\varrho_{m}^{\rm th}$ evolves into
\begin{equation}
\label{final1}
\hat{\cal U}_t(\ket{+}_a\!\bra{+}\varrho_{m}^{\rm th}\,)\hat{\cal U}^\dag_t\!=\!\int\!{d}^2\alpha{\cal P}(\alpha,V)\ket{\psi(t)}\!\bra{\psi(t)},
\end{equation}
which reduces to the pure case of Eq.~(\ref{catstate}) for $T=0$. We proceed to show that the coupling mechanism described above is characterized by interesting features, at the core of current experimental and theoretical interests~\cite{demartini,jeong1,jeongralph}. Let us consider the case of $\phi=\pi/2$, $V=1$ (i.e. $T=0$) and $\alpha\in\mathbb{R}$, which gives $\ket{\psi(\tau)}(\ket{1,\alpha}+\ket{0,\alpha-\eta t})/\sqrt2$. This entangled state represents a mesoscopic instance of a pure Schr\"odinger-cat state. Interestingly, it has been discussed that a faithful implementation of the Schr\"o{d}inger's cat paradox would use a mesoscopic subsystem initially prepared in a thermal state, rather than a pure one~\cite{demartini,jeong1,jeongralph}. The state in Eq.~(\ref{final1}) is a significant example of such case. Unravelling the entanglement properties of this state is demanding due to the difficulty of finding an analytical tool for its undisputed revelation. In order to gain insight, here we propose to follow two paths. 

The first relies on the nonlocality properties of this class of states, induced by the strong entanglement between the subsystems. Following Ref.~\cite{banaszek,banwod}, the microscopic part is projected along the direction ${\bf n}=(\sin\theta,0,\cos\theta)$ of the single-qubit Bloch sphere while the mesoscopic one is probed by using the displaced parity observable $\hat{\Pi}(\beta)=\hat{D}^\dag(\beta)(-1)^{\hat{b}^\dag\hat{b}}\hat{D}(\beta)$, where $\hat{D}(\beta)$ is the displacement operator of amplitude $\beta=\beta_r+i\beta_i$. This approach has been used recently to address the micro-macro non-locality in an all-optical setting~\cite{spagnolo2011}. The correlation function for a joint measurement is thus 
\begin{equation}
{\cal C}(\beta,\theta)=\int{d}^2\alpha{\cal P}(\alpha,V)\bra{\psi(t)}({\bf n}\cdot\hat{{\bm \sigma}})\otimes\hat{\Pi}(\beta)\ket{\psi (t)}
\end{equation}
 and a Bell-Clauser-Horne-Shimony-Holt (Bell-CHSH) inequality is formulated as $|{\cal C}(0,\theta_1)+{\cal C}(0,\theta)+{\cal C}(\beta,\theta_1)-{\cal C}(\beta,\theta)|\le{2}$~\cite{chsh}. Any state satisfying this constraint can be described by a local-realistic theory. Let us first discuss the pure case of $V=1$, which gives  
\begin{equation} 
\begin{split}
{\cal C}(\beta,\theta)&=\frac{1}{2}e^{-2(d^2+\eta^2 t^2+|\beta|^2+\beta_r\eta t-2\beta_rd)}\\ 
&\times[\cos\theta(e^{4d\eta t-2\eta t \beta_r}\!-\!e^{2\eta^2 t^2+2\eta t \beta_r})\!\\ &+\!2e^{\eta t(2{d}+\frac{3}{2}\eta t)}\cos({2\eta t \beta_i})\sin\theta].
\end{split}
\end{equation}
At $\eta t=0$, the microscopic and mesoscopic subsystems are uncorrelated and ${\cal C}(\beta,\theta)$ can indeed be factorized. For a set value of $d$ and a non-zero value of $\eta t$, we observe violation of the Bell-CHSH inequality as illustrated in Fig.~\ref{CHSHoneMirr}. Moreover,  there is a range of values of $\theta$ ($\sim\pi/2$) where, for $d\neq{0}$, the local-realistic bound is violated, symmetrically with respect to $d=0$. When the thermal character of the mesoscopic part is considered, the expression for the correlation function becomes cumbersome and we omit it. However, {\it the strong entanglement between microscopic and mesoscopic subsystems allows violation of Bell-CHSH inequality also in the mixed-state case}: the dotted curve in Fig.~\ref{CHSHoneMirr} corresponds to $V\simeq{5}$.  Beyond this value, the inequality is no longer violated. 

The second path we follow uses the technique put forward in Ref.~\cite{bosekim} and later reprised by Ferreira {\it et al.} in Ref.~\cite{ferreira}. In this approach, Eq.~(\ref{final1}) is projected onto a bidimensional subspace spanned by the microscopic states $\{\ket{0},\ket{1}\}_a$ and the phononic ones $\{\ket{p},\ket{p+1}\}_m$ ($p\in\mathbb{Z}$). The entanglement within Eq.~(\ref{final1}) cannot be increased by this projection, which is just a local operation. Thus, by quantifying the entanglement for fixed $p$, we provide a lower bound to the 
overall quantum correlations in the state of the system. As a measure for entanglement in each $2\times{2}$ subspace we use the {\it logarithmic negativity}, which accounts for the degree of violation of the positivity of partial transposition criterion~\cite{peres,horo3,lee,plenio05}. An example of the results achieved with this method is given in the inset of Fig.~\ref{CHSHoneMirr}, where we show the case of $d=2$ and $p=0,1,2$. Entanglement is found in each subspace with fixed $p$, up to values of $V\sim{5}$, strengthening our findings about the resilience of non-classical correlations set by the coupling being studied.

\subsection{Non-classicality of the mirror}
We now consider the effects of the microscopic-mesoscopic interaction over the state of the movable mirror. This is a hot topic in the current research of opto and electro-mechanical systems. The grounding of opto/electro-mechanical devices as potential candidates for quantum information processing requires the design of protocols for the preparation of non-classical states of massive mechanical systems. Various attempts have been performed in this direction, mainly at the nano-scale level, where a cantilever can be capacitively coupled to a superconducting two-level system~\cite{armour,rabl,tian,rodrigues}. 
\begin{figure}[b] 
\centerline{\scalebox{0.25}{\includegraphics{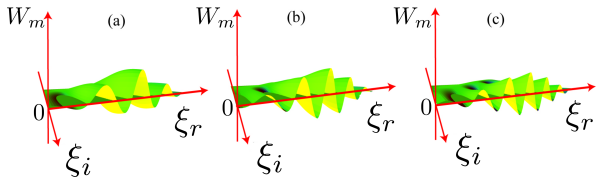}}}
\caption{(Color online) Wigner function of the conditional mirror state against $\xi_r=\text{Re}(\xi)$ and $\xi_i=\text{Im}(\xi)$, for $V=3$ and $d=0$. Panels {\bf (a)}, {\bf (b)}, {\bf (c)} correspond to $\eta t=2,3,4$ respectively. }
\label{Wignerevolution}
\end{figure}

Let us consider the case of $\phi=0$. The optomechanical evolution encompassed by $\hat{\cal U}_t$ alone is unable to give rise to any non-classicality in the state of the mirror. This is easy to check simply by tracing out the state of the atom in Eq.~(\ref{catstate}), which would leave us with a statistical mixture of two displaced mirror's states. On the other hand, a conditional process is able to project the coherence of a quantum mechanical superposition and simultaneously get rid of the atomic degree of freedom \cite{KochPRL,RitterNature,SpechtNature,MonteiroNJP,Kiesel,MauroPRA}. In order to illustrate our claim, we consider an initial state of the system having the form $\rho(0){=}|\varphi\rangle \langle \varphi |{\otimes}\rho_m(0)$ where $|\varphi\rangle{=}c_0 |0\rangle{+}c_1 |1\rangle$  is a pure state of the atom and $\rho_m(0)$ is an arbitrary state of the mechanical mode. We then project the atomic part of the evolved state $\hat{\cal U}_t \rho(0)\hat{\cal U}_t^\dag$ onto $|\varphi\rangle \langle \varphi |$, thus post-selecting the mechanical state $\rho_m(t) =  \langle \varphi | \hat{\cal U}_t |\varphi \rangle \rho_m(0)\langle \varphi | \hat{\cal U}_t^\dag |\varphi\rangle$. Therefore, the state of the mirror undergoes an effective evolution driven by the operator 
	\begin{equation}
\langle \varphi | \hat{\cal U}_t | \varphi \rangle = |c_1|^2 \hat\openone + |c_0|^2\hat{D}(-i \eta t). 
\label{pUpOneMirr}
	\end{equation}
In the remainder of this paper, we consider again the case where $|{\varphi}\rangle=|{+}\rangle\equiv(|0\rangle + |1\rangle)/\sqrt{2}$, which optimizes the performance of our scheme  terms of the degree of non-classicality enforced in the mechanical subsystem. For an initial coherent state of the mirror, i.e. $\rho_m(0)=|\alpha\rangle\langle\alpha|,$ applying the conditional time evolution operator in Eq.~(\ref{pUpOneMirr}) leads to $\ket{\mu_+}_m={\cal N}_+(\ket{\alpha}+{e}^{-i\varPhi(t)}\ket{\alpha-i\eta t})_m$, where ${\cal N}_+$ is the normalization factor. Depending on the value of $\eta t$, such states exhibit quantum coherences. Obviously, the thermal convolution inherent in the preparation of mirror's state $\varrho_m$ may {\it blur} them. In what follows we prove that this is not the case for quite a wide range of values of $V$. 

\begin{figure}[t]
\includegraphics[width=.8\columnwidth]{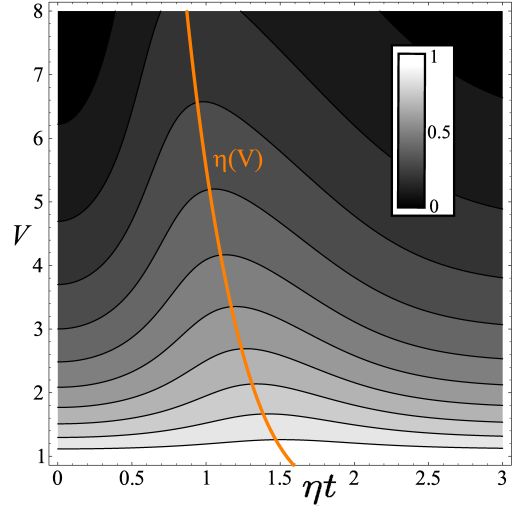}
\caption{(Color online) Density plot of fidelity against $V$ and $\eta$. Darker regions correspond to smaller values of $F_{W}$. }
\label{fidelity}
\end{figure}
The figure of merit that we use to estimate non-classicality is the negativity in the Wigner function associated with the mirror state resulting from the measurement performed over the atomic part of the system.
The Wigner function for a single bosonic mode is  defined as 
\begin{equation}
{W(\mu)=\frac{1}{\pi}\int d^2\nu e^{\mu \nu^* - \mu^* \nu}  \chi(\nu)},
\label{WignerDef}
\end{equation}
where $\mu {\in}\mathbb{C}$ and $\chi(\nu){=}\text{Tr}[\hat{D}(\nu)\rho]$ is the Weyl characteristic function.  Considering an initial thermal state of the mirror and applying the conditional unitary evolution operator given in Eq.~(\ref{pUpOneMirr}), the Wigner function of the mirror after the post-selection process is 
\begin{equation}
\begin{split}
W_m(\mu)=&{\cal M}^{-1}e^{-\frac{2|\mu|^2+2\eta t \mu_i+\eta^2 t^2}{V}}\\ \times&[\cosh(\frac{\eta^2 t^2+2\eta t\mu_i}{V})+e^{\frac{\eta^2 t^2}{2V}}\cos(2\eta t\mu_r)]
\end{split}
\end{equation}
 with ${\cal M}=(1+e^{-\frac{V\eta^2}{2}})\pi{V}/2$. The behavior of $W_m(\mu)$ in the phase space is shown in Fig.~\ref{Wignerevolution}, where we clearly see the appearance of regions of negativity, witnessing non-classicality of the corresponding state as induced by our microscopic-to-mesoscopic coupling. Interference fringes are created between two positive Gaussian peaks (not shown in the figure) corresponding to the position, in the phase space, of mutually displaced coherent states. This reminds of the Wigner function of a pure Schr\"odinger cat state although, as we see later, the analogy cannot be pushed. Remarkably, in contrast with the fragility of the nonlocality properties of the microscopic-mesoscopic state, $W_{m}(\mu)$ has a negative peak of $-0.01$ up to $V\sim100$, which implies strong thermal nature of the mirror state. For a mechanical system embodying one of the mirrors of a cavity, $\omega_m/2\pi\sim5$MHz is realistic~\cite{parameters}. For $V=10$ ($100$), this corresponds to an effective temperature of $1$mK ($10$mK), {\it i.e.} energies $10$ ($100$) times larger than the ground-state energy of the mirror.

It is interesting to compare the mixed state resulting from the thermal convolution to a pure state in Eq.~(\ref{catstate}) (with $\phi=0$). As a measure of the closeness of two states, we use quantum fidelity between a mixed and a pure state written as the overlap between the corresponding Wigner functions $F_W=\pi\int{d}^2\mu{W}_P(\mu)W_M(\mu)$, where $W_P(\mu)$ ($W_M(\mu)$) is the Wigner function of the pure (mixed) state. $F_W$ is shown in Fig.~\ref{fidelity} against $\eta\tau$ and $V$. While the thermal effect reduces the value of the fidelity as $V$ grows, the behavior of $F_W$ against $\eta t$ is, surprisingly, non-monotonic. At a given $V$, there is always a finite value of $\eta t$ associated with a maximum of $F_W$. Remarkably, the values of $\eta t$ maximizing $F_W$ differ from those at which the Wigner function achieves its most negative value. 

\begin{figure}[t]
\includegraphics[width=.9\columnwidth]{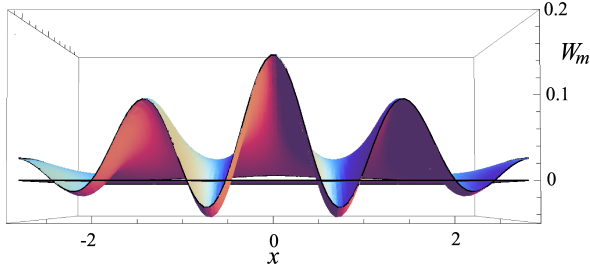}
\caption{(Color online) Wigner function of the mirror under dissipation, for $\gamma\sim{0.1}\eta$ and $V=5$.}
\label{WdissOneMirr}
\end{figure}
\subsection{Finite temperature dissipative dynamics}
\label{subsectionOMDissipative}
So far, we have assumed a movable mirror of large mechanical quality factor. The progresses recently accomplished in fabrication processes guarantee very small mechanical dissipation rates. However, they are not yet negligible and their effect should be considered in any proposal for quantumness in optomechanical devices. We thus include mechanical losses in our analysis, looking for their effects onto the non-classicality induced in the movable mirror. We concentrate on the finite-temperature dissipative mechanism described by 
\begin{equation}
{\cal L}^{V}(\rho)=\frac{\gamma}{2}\big[(2\hat{b}\rho\hat{b}^\dag-\{\hat{b}^\dag\hat{b},\rho\})+(V-1)(\hat{b}\rho-\rho\hat{b},\hat{b}^\dag)\big], 
\end{equation}
which is the weak-damping limit of the Brownian-motion master equation~\cite{walls}. The density matrix $\rho$ describes the state of the atom-mirror system. The full master equation, including the unitary part $-i[\hat{\cal H}_e,\rho]$, is easily translated into a set of equations of motion for the mirror reduced density matrix obtained by considering the projections onto the relevant atomic states $\rho_{ij}={}_a\langle{i}|\rho|j\rangle_a~(i,j=0,1)$. These can then be recast as Fokker-Planck equations for the Wigner functions $W_{ij}$ of such mirror's state components. These read 
\begin{equation}
\partial_t{\bf W}(x,p, t)={\bf M}{\bf W}(x,p, t)+\tilde{\cal L}_d{\bf W}(x,p, t),
\label{FokkerPlanck}
\end{equation}
where  
\begin{equation}
\begin{split}
&{\bf W}(x,p, t){=}\left[\begin{matrix}W_{00}(x,p,t)\\W_{01}(x,p,t)\\W_{10}(x,p,t)\\W_{11}(x,p,t)\end{matrix}\right],\\
&{\bf M}{=}\sqrt 2\eta\left[\begin{matrix}\partial_p&0&0&0\\ 0&-\frac{ix+\partial_p}{2}&0&0\\0&0& \frac{ix+\partial_p}{2}&0\\0&0&0& 0\end{matrix}\right],\\
&\tilde{\cal {L}}_d=\Big[\frac{\gamma}{2}(x\partial_x+p\partial_p)+\frac{\gamma}{4}V(\partial^2_{p^2}+\partial^2_{x^2})+\gamma\Big]\openone,
\end{split}
\end{equation}
where we have introduced the quadrature variables $x=\sqrt 2\text{Re}(\mu), p=\sqrt 2\text{Im}(\mu)$. Each of these equations preserves the Gaussian nature of the corresponding Wigner function's component, whose time-evolved form is taken from the ansatz 
\begin{equation}
W_{ij}(x,p,t)\propto[{\text{det}}({\bf D}_{ij})]^{-1/2}{e^{-\frac{1}{2}{{\bf q}^T_{ij}{\bf D}^{-1}_{ij}{\bf q}_{ij}}+i\Theta_{ij}(t)}}
\label{ansatz}
\end{equation}
with 
\begin{equation}
{\bf q}_{ij}=\left[\begin{matrix}x-\overline{x}_{ij}\\p-\overline{p}_{ij}\end{matrix}\right],~{\bf D}_{ij}=\left[\begin{matrix}\sigma^x_{ij}&\sigma^{xp}_{ij}\\\sigma^{xp}_{ij}&\sigma^{p}_{ij}\end{matrix}\right]
\label{qandcovariance}
\end{equation}
 parameterized by the time-dependent mean values $\overline{x}_{ij},\overline{p}_{ij}$ and variances $\sigma^{x,p,xp}_{ij}$ of the variables $x,p$ and $xp$. We have also introduced the time-dependent phases $\Theta_{ij}$'s which account for the contributions from $\varPhi(t)$ in Eq.~(\ref{catstate}). The solution is readily found to be $\sum_{i,j=0,1}W_{ij}(x,p,t)$ (apart from the normalization factor), which gives back the non-Gaussian character of the mirror's state. The negativity of the Wigner function can be studied at set values of $\gamma$ and $T$ and chosing the time at which the ideal case achieves the most negative value. The results are shown in Fig.~\ref{WdissOneMirr}, where we see that non-classicality is found even for quite a large value of $\gamma/\eta$. Clearly, this results from a subtle trade off between temperature and mechanical quality factor. Although small $\gamma$ and $T$ guarantee non-classicality, such a behavior is still present at $\gamma/\eta\sim{0.1}$ and for $T$ well above the ground-state one.

\section{Two Mirrors }\label{TM}
In this section, we will consider a different setup, where both cavity mirrors are free to oscillate around their equilibrium positions and they are both interacting with a three level atom inside the cavity. Using this setup, we can study the correlations between the two mesoscopic systems and their quantum features. In this section, we will only focus on the conditional evolution of the two mirrors after a measurement of the atomic subsystem. 

\subsection{Hamiltonian and conditional unitary evolution}
\label{subsectionTMH}

Let us  consider the same  Fabry-Perot cavity discussed in Sec \ref{subsectionOMModel}, pumped by a  laser field at frequency $\omega_p$ and with a  three-level $\Lambda$-type atom  trapped within the mode-volume of the cavity field. The  model is very similar to the one describing the single mirror case, with the difference that  here the two mirrors of the cavity are both able to oscillate around their equilibrium positions and they are modeled as two harmonic oscillators with frequencies $\omega_1$ and $\omega_2.$  By moving to an interaction picture respect to the same operator considered  in the one-mirror scheme, the Hamiltonian of the system can be written in the same form as the one given in Eq.~(\ref{HTot}), where only the terms involving the mirror's degrees of freedom are changed to take into account the addition of the second mirror. These terms read 
\begin{equation}
\begin{split}
&\hat{\cal H}_{\rm m}{=}\sum^2_{j=1}\omega_j \hat{b}^\dag_j \hat{b}_j,~~\hat{\cal H}_{\rm mc}{=}\hat{a}^\dag \hat{a} \sum^2_{j=1}(-1)^{j-1}\chi_j (\hat{b}^\dag_j{+}\hat{b}_j),
\end{split}
\end{equation}
where the bosonic operators $\hat{a}^\dag,\hat{a}$ and $\hat{b}^\dag_j,\hat{b}_j$ refer to the cavity field and the two mechanical mirrors, respectively. By assuming a large cavity quality factor and a small spontaneous emission rate from $\ket{e}$, in the limit of $(\Delta,\delta){\gg}(\Omega, g)$ we can eliminate both the cavity field and the excited atomic level, thus arriving at the effective atom-mirrors Hamiltonian 
\begin{equation}
\hat{\cal H}_{\rm eff} = |0\rangle \langle 0|\otimes \sum^2_{j=1}(-1)^{j-1}\eta_j (\hat{b}^\dag_j{+}\hat{b}_j)
\end{equation}
with $\eta_j= ( \Omega^2 g^2/ \delta^2 \Delta^2)\chi_j. $ The corresponding time-evolution operator is $\hat{\cal U}_t = |1\rangle \langle 1| + |0\rangle \langle 0|\otimes \hat{D}_1(-i \eta_1 t)\otimes \hat{D}_2(i \eta_2 t)$, where $\hat{D}_j(\zeta)  = \exp[\zeta \hat{b}^\dag_j -\zeta^* \hat{b}_j]$ is the displacement operator for mode ${j=1,2}$ \cite{walls}. 
In analogy with the one-mirror case,  the resulting dynamics of the mechanical systems is thus a conditional displacement controlled by the state of the atomic part: while nothing happens to the mechanical modes when the atom is prepared in $\ket{1}$, their state gets displaced in phase space when the atomic state is $\ket{0}$. In what follows we generalize the analysis performed in the previous Section and show how this mechanism, complemented with an appropriate post-selective step, results in non-classicality of the mechanical subsystem.

The generalization of the conditional time evolution operator given in Eq.~(\ref{pUpOneMirr}) to the two-mirrors case is straightforward. The new operator simply reads
\begin{equation}
\langle \varphi | \hat{\cal U}_t | \varphi \rangle = |c_1|^2 \hat\openone + |c_0|^2\hat{D}_1(-i \eta_1 t)\hat{D}_2(i \eta_2 t) \label{pUp}
\end{equation}
with $\hat\openone$ the identity operator. We consider again the case in which $| \varphi \rangle = |+\rangle $ and  the initial state of the mirror is  $\rho_m(0)=|\alpha_1, \alpha_2\rangle\langle\alpha_1,\alpha_2|$ where $|\alpha_j\rangle$ is a coherent state of mode $j$ having amplitude $\alpha{\in}\mathbb{C}$. The state of the mirrors at time $t$ is
	\begin{equation}
|\psi_m(t)\rangle =(| \alpha_1, \alpha_2 \rangle  + e^{-i \varPhi(t)}  | \beta_1(t) , \beta_2(t) \rangle)/{\sqrt2}
\label{psit}
	\end{equation}
where $\varPhi(t)=\sum^2_{j=1}(-1)^{j-1}\eta_j\text{Re}\{\alpha_j\}t$ and $\beta_j(t)=\alpha_j{+}(-1)^{j} i \eta_j t~(j{=}1,2)$. Eq.~(\ref{psit}) is an Entangled Coherent State (ECS) of modes $1$ and $2$~\cite{sanders}. Its Von Neumann entropy depends on a delicate trade off among the amplitudes $\alpha_j(t)$ and $\beta_j(t)$. 
ECSs play an important role in continuous-variable (CV) quantum information processing as a valuable resource for communication and computation~\cite{jeong2}.

\subsection{Mirror-Mirror correlations}
\begin{figure}[t]\centering
\includegraphics[width=.8\columnwidth]{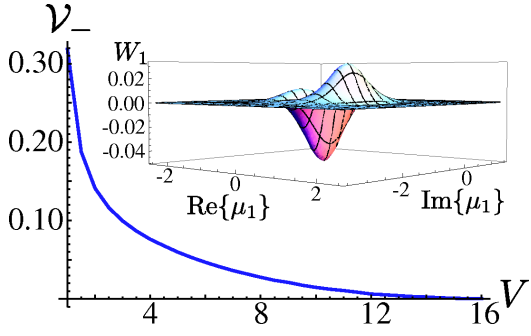}
\caption{(Color online) Negative volume of $W(\mu_1,\mu_2)$  against $V$ for $\eta{t}=5$. Inset:  Wigner function $W_1(\mu_1)$ at $\mu_2{=}-(1+i)$, $\eta{t}{=}2$ and $T{=}0$. }\label{WignerT}
\end{figure}
In Sec \ref{subsectionTMH} we have considered the simple case  in which  the two oscillators are initialized in a pure coherent state. This example is instructive and, as we will see later, mathematically  useful.  However, as pointed out in Sec. \ref{subsecOMEnt}, the interaction of the two oscillators with the thermal bath has to be taken into account, and it is realistic to assume a initial thermal state for the two mirros. The thermal state of a single bosonic mode is given by Eq.~(\ref{thermalstate}) . In the case of two modes,  the initial mechanical state is ${\rho_{\rm m}(0){=}\varrho^{\rm th}_1{\otimes}\varrho^{\rm th}_2},$ and it evolves under the action of $\langle+|\hat{\cal U}_t|+\rangle$ so as to give 
\begin{equation}
\rho_m(t){=}\int d^2 \alpha_1 d^2 \alpha_2 {\cal P}(\alpha_1,V){\cal P}(\alpha_2,V)|\psi_m(t) \rangle \langle \psi_m(t)|.
\label{EvolTherStateTwoMirr}
\end{equation} 
As in the one mirror setup, we now show that, despite the thermal convolution at the basis of the definition of $\rho_{}(t),$ the mechanical state of two mirrors can exhibit strong non-classical features even at non-zero temperature. We will focus on two different signatures of non-classicality: the negative values of the Wigner function associated with the state $\rho_m(t)$ and the non-local correlations between the two mirrors. The Wigner function of a two-modes system is defined as the straightforward generalization of Eq.~(\ref{WignerDef}), i.e. 
\begin{equation}
{W(\mu_1,\mu_2)=\frac{1}{\pi^{2}}\!\int d^2\nu_1 d^2\nu_2 \prod^2_{j=1}e^{\mu_j \nu^*_j - \mu^*_j \nu_j}  \chi(\nu_1,\nu_2)}
\end{equation}
where $(\mu_j,\nu_j){\in}\mathbb{C}$ and $\chi(\nu_1,\nu_2){=}\text{Tr}[\hat{D}_1(\nu_1)\hat{D}_2(\nu_2)\rho]$ is the two-modes Weyl characteristic function.
Together with the study of Wigner function's negativity, we also investigate the quantum correlations between the two mirrors. To overcome the problem of inferring non-classical correlations in a mixed non-Gaussian state of a CV system, which  is a very demanding task due to the lack of appropriate entanglement measures, we use the same approach taken in the previous Section, which relies on the investigation of Bell inequality violations. This route is particularly viable in our case as we can take advantage of the dualism between density matrix and Wigner function for CV states. Here,  one can formulate a Bell-CHSH test using the two-mode Wigner function associated to $\rho_{m}(t)$.

\begin{figure}[b]\centering
\includegraphics[width=.8\columnwidth]{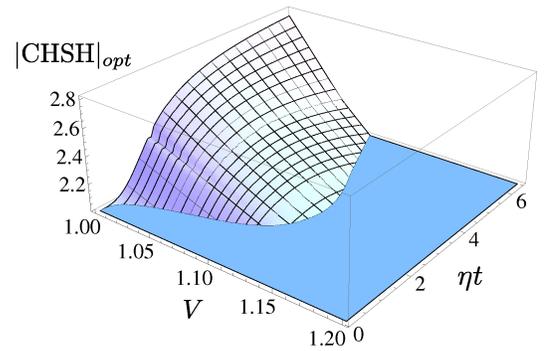}
\caption{(Color online) Numerically optimized violation of the Bell-CHSH inequality for the two-mirror state against $\eta{t}$ and $V$. }\label{WignerTCHSH}
\end{figure}
To begin with, one can study the behavior of the single-mirror Wigner functions calculated for a fixed point $\mu_0$ in the other mirror phase space, i.e.  $W_j(\mu_j){=}W(\mu_j,\mu_i = \mu_0)$ with $i{\neq}{j}{=}1,2$.  It is seen from the inset of  Fig.~\ref{WignerT} that, depending on the operating conditions of the system, $W_1(\mu_1)$ [equivalently $W_2(\mu_2)$] can be considerably negative, thereby proving its non-classical nature. This is remarkable, especially when compared to the case of a standard optomechanical setting where a mechanical mirror is coupled to the field of an optical resonator. There, in fact, it can be proven that the state of the mechanical subsystem is only {\it classically squeezed}  and the device cannot be utilized in order to engineer non-classical states of the movable mirror~\cite{mauro2}. Differently, using the mechanism we propose here, we have checked that the negative regions of $W_1(\mu_1)$ persist even at non-zero temperature. These considerations can be strengthened by extending them to the Wigner function of both the mechanical mirrors and studying the {\it negative volume} ${\cal V}_-{=}\int{d}^2\mu_1{d}^2\mu_2[|W(\mu_1,\mu_2)|{-}W(\mu_1,\mu_2)]/2$. In Fig.~\ref{WignerT}, ${\cal V}_-$ is plotted against $V$ for ${\eta{t}=5}$, revealing that non-classicality persists up to $V{\sim}{10}$, {\it i.e.} well above zero temperature. We give an estimate of actual temperatures corresponding to such order of magnitude for $V$ later on.

We now pass to the study of the Bell-CHSH inequality test~\cite{chsh} to infer non-classical correlations shared by the mechanical systems. For a two-mode bosonic system, the Bell-CHSH inequality can be re-cast in terms of the expectation values $\langle\hat\Pi_1(\mu_1)\otimes\hat\Pi_2(\mu_2) \rangle$, with the displaced parity operator $\hat{\Pi}_i(\mu_j) =\hat{D}_j(\mu_j)(-1)^{\hat{b}^\dag_j\hat{b}_j }\hat{D}_j^\dag(\mu_j)$, as before~\cite{banaszek}, in terms of which $W(\mu_1,\mu_2)=(4/\pi^2)\langle\hat\Pi_1(\mu_1){\otimes}\hat\Pi_2(\mu_2) \rangle$. The CHSH function can thus be written as
\begin{equation}
\text{CHSH}{=}\frac{\pi^2}{4}[W(\mu_1,\mu_2){+}W(\mu'_1,\mu_2){+}W(\mu_1,\mu'_2){-}W(\mu'_1,\mu'_2)].
\end{equation}
Any local realistic theory imposes the bound $|\text{CHSH}|\le{2}$. If the mechanical state is such that $|\text{CHSH}|>2$, correlations of non-classical nature are necessarily shared by the two mirrors. In Fig.~\ref{WignerTCHSH}  we show that, although  hindered by the thermal nature of the mechanical modes, the two-mirror state violates the local realistic bound up to $V=1.1$, which corresponds to ${T \approx0.1}$mK (5$\mu$K) at ${\omega_m/2\pi\sim6}$MHz (${300}$KHz), a frequency  easily  achievable by current experimental setups~\cite{Groeblacher}. This shows that the mechanical state remains non-classically correlated even for thermal energies that are 10 times larger than the ground-state energy of each mirror~\cite{commentT0}.  

The decreasing behavior of the CHSH function at ${T{>}0}$ can be explained by considering that, under such conditions, the coherences in the two-mirror state are suppressed. In fact, let us study the off-diagonal terms of $\rho_m(t)$ in the coherent-state basis. These are given by 
$\int d^2 \alpha_1 d^2 \alpha_2 P(\alpha_1,V)P(\alpha_2,V)
e^{i\varPhi(t)}|\alpha_1, \alpha_2 \rangle \langle \beta_1(t), \beta_2(t) |$ 
and Hermitian conjugate. As a function of $\text{Re}(\alpha_j)$, the phase factor $e^{i\varPhi(t)}$ oscillates at frequency $\eta_j t$. At $T{=}0$, $P(\alpha_j,1)$ becomes a bidimensional Dirac delta-function $\delta^2(\alpha_j)$, which sets the phase factor to unity. At the same time, by increasing $\eta_j{t}$, the components of the ECS entering state $\rho_{m}(t)$ become increasingly orthogonal, which optimizes the violation of the CHSH inequality. Differently, at finite temperature $P(\alpha_j,V)$ has a non-null width within which the increasingly oscillating time-dependent phase factor is eventually averaged to zero. This occurs more rapidly as $V$ grows.

\subsection{Dissipative dynamics}

We now proceed to include the mechanical damping of the two oscillator in our analysis on the same lines followed in Sec \ref{subsectionOMDissipative}. We consider  the dynamics of the mirror-atom density matrix $\rho$ as driven by the weak-damping limit of the standard Brownian-motion superoperator, whose generalization to the a two mirrors system reads as 
\begin{equation}
\hat{\cal L}^V(\rho)=\sum_{j{=}1,2}\frac{\gamma}{2}(2\hat{b}^{}_j\rho\hat{b}^\dag_j{-}\{\hat{b}_j^\dag\hat{b}^{}_j,\rho\}{+}(V{-}1)[\hat{b}^{}_j\rho{-}\rho\hat{b}^{}_j,\hat{b}_j^\dag]).
\end{equation} 
From such master equation one can obtain with standard techniques four Fokker-Planck equations for the Wigner functions $W_{ij}$ of the mechanical state components associated  the atomic operator $\ket{i}\bra{j}$ ($i,j{=}0,1$). The Foller-Planck equations can be written in the same form given in Eq.~(\ref{FokkerPlanck}), and each equation is solved by using the Gaussian ansatz in Eq.~(\ref{ansatz}), which is worth recalling 
\begin{equation}
W_{ij}(x,p,t){\propto}[{\text{det}}({\bf D}_{ij})]^{-1/2}{e^{{-\frac{1}{2}{{\bf q}^T_{ij}{\bf D}^{-1}_{ij}{\bf q}_{ij}}+i\Theta_{ij}(t)}}}
\end{equation}
Here the vector ${\bf q}$ and the covariance matrix ${\bf D}_{ij}$  are the generalization of Eq.~(\ref{qandcovariance}) to the two-mode case and are given by  
\begin{equation}
\begin{split}
{\bf q}_{ij}=\left[\begin{matrix} x_1{-}\overline{x}_{1,ij}\\p_1{-}\overline{p}_{1,ij}\\x_2{-}\overline{x}_{2,ij}\\p_2{-}\overline{p}_{2,ij}\end{matrix}\right], {\bf D}_{ij}=\left[\begin{matrix}\sigma^{x_1x_1}_{ij}&\sigma^{p_1x_1}_{ij}&\sigma^{x_2x_1}_{ij}&\sigma^{p_2x_1}_{ij}\\ \sigma^{x_1p_1}_{ij}&\sigma^{p_1p_1}_{ij}&\sigma^{x_2p_1}_{ij}&\sigma^{p_2p_1}_{ij}\\ \sigma^{x_1x_2}_{ij}&\sigma^{p_1x_2}_{ij}&\sigma^{x_2x_2}_{ij}&\sigma^{p_2x_2}_{ij}\\ \sigma^{x_1p_2}_{ij}&\sigma^{p_1p_2}_{ij}&\sigma^{x_2p_2}_{ij}&\sigma^{p_2p_2}_{ij}\end{matrix}\right].
\end{split}
\end{equation}
As explained  in the previous section, the sum of the four term $\sum_{i,j=0,1}W_{ij}(x,p,\tau)$ gives the full non-gaussian solution of the Fokker-Planck equations, and  the negativity of the Wigner function can be use to witness non-classicality.
\begin{figure}[b]\centering
{\bf (a)}\hskip4cm{\bf (b)}\\    
\includegraphics[width=.45\columnwidth]{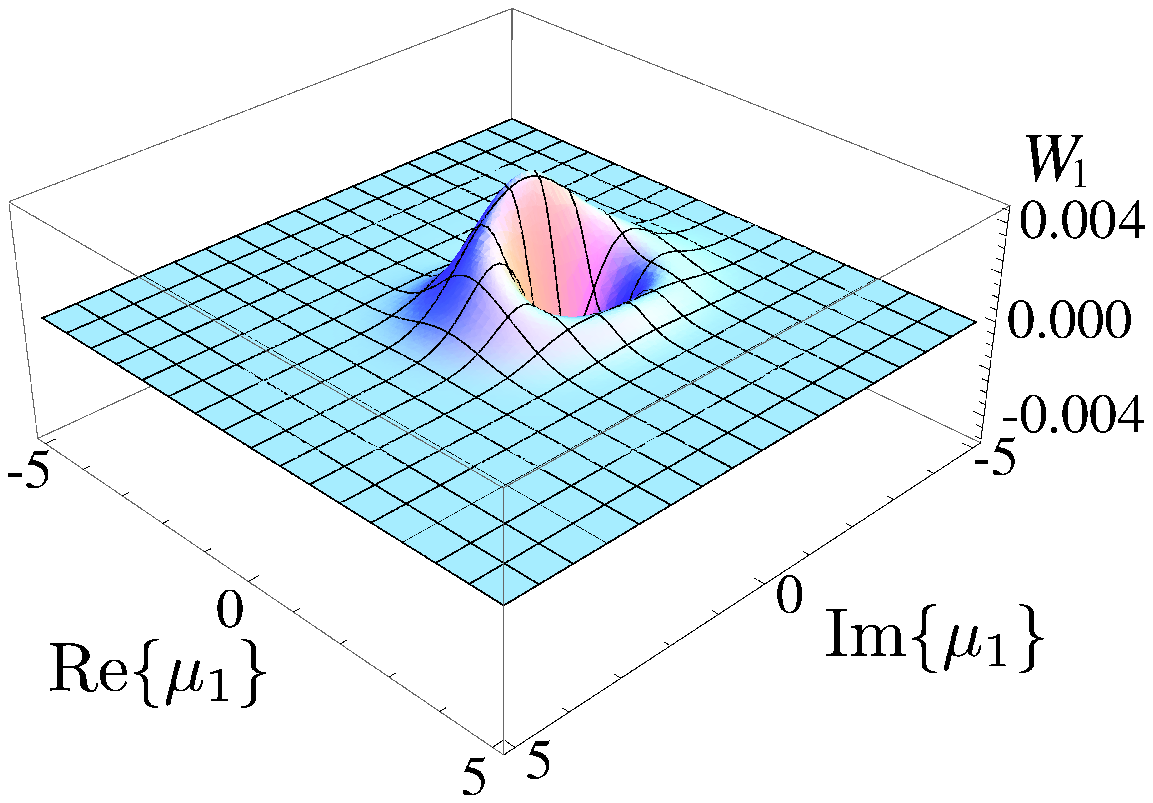}~\includegraphics[width=.45\columnwidth]{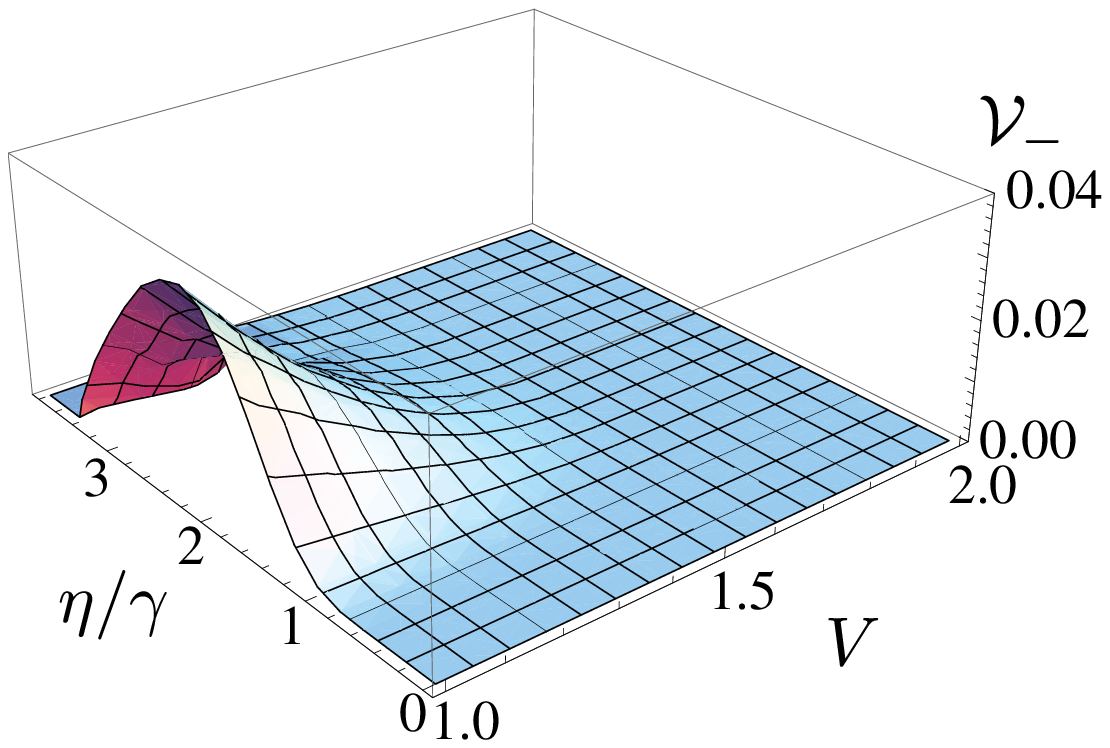}\caption{(Color online) Wigner function for a mechanical system open to dissipation. {\bf (a)} Wigner function of a single mirror for $\mu_2{=}1{+}i$, $\eta/\gamma{=}2$, $\gamma{t}{=}V{=}1$. {\bf (b)} ${\cal V}_-$ against $V$ and $\eta/\gamma$ for $\gamma{t}=1$ (we assume that all the relevant parameter are the same for both mirrors). }\label{NoisyWigner}
\end{figure}
\begin{figure}[t]\centering  
\includegraphics[width=.8\columnwidth]{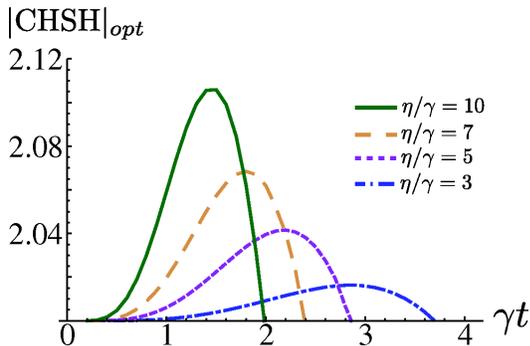}
\caption{(Color online) Violation of the CHSH inequality as a function of $\gamma t$ for four values of $\eta/\gamma$.}\label{NoisyWignerCHSH}
\end{figure}
Fig.~\ref{NoisyWigner}{\bf(a)} reveals that $W(\mu_1,\mu_2)$ exhibits considerable regions of negativity also for $\gamma{\neq}0$. As expected, the negativity of the Wigner function increases when the coupling constant $\eta$ becomes larger than the damping rate. In this situation it is indeed possible to neglect the dissipation of the mirror and recover the purely unitary dynamics treated above. Interestingly, the Wigner function has still negative values when $\eta \sim \gamma,$ which means that in the dissipative regime the state of the two mirrors is non-classical. The decrease of ${\cal V}_-$ as $\eta/\gamma\gg{1}$ shown in Fig.~\ref{NoisyWigner}{\bf(b)} is simply due to our choice for the interaction time. By adjusting $t$, non-zero values of ${\cal V}_-$ are retrieved. The interplay between $\gamma,\eta$ and $t$ in setting non-classicality in the mechanical state can be best seen by studying non-locality. As shown in Fig.~\ref{NoisyWignerCHSH}, as $\eta/\gamma$ increases for damped mechanical systems at zero-temperature, the interaction-time window has to be set so as to maximize the degree of violation of the CHSH inequality. As expected, the violation increases with the ratio between the coupling constant and the decay rate. However, large values of $\eta/\gamma$ correspond to shorter time-windows for the violation to occur. This point can be understood solving explicitly the open-system dynamics corresponding to a low-temperature bath in an alternative way. 

Following the approach used in \cite{noiGP}, we divide the evolution time as $t{=}N\delta{t}$, with $\delta t/t{\ll}{1}$ and approximate the dynamics of the total system as a sequence of the unitary dynamics ruled by $\hat{U}_t$ and a purely dissipative one. After $N$ steps, the evolved state reads 
\begin{equation}
\rho(N \delta t){=}\big(\hat{\mathcal{D}}^1_{\delta t}\hat{\mathcal{D}}^2_{\delta t}\hat{\mathcal{U}}_{\delta t}\big)^N \rho(0)
\end{equation}
where we have introduced the superoperators 
\begin{equation}
\begin{split}
&\hat{\mathcal{D}}_{\delta t}^{j} \rho{=} e^{{\hat{\mathcal{L}}^{V{=}1}_j  \delta t} }\rho,~~\hat{\mathcal{U}}_{\delta t}  \rho{=}\hat{U}_{\delta t} \rho \hat{U}^\dag_{\delta t}
\end{split}
\end{equation}
and where ${\rho(0){=}|+\rangle \langle +|{\otimes}|\alpha_1,\alpha_2\rangle \langle \alpha_1,\alpha_2|}$ is the initial state. This approach is particularly useful in treating a damped harmonic oscillator. Indeed, the action of the dissipative superoperator $\hat{\mathcal{D}}_{\delta{t}}^{j}$ on the diadic form $|\lambda\rangle\langle\sigma|$ (with $\ket{\lambda}$ and $\ket{\sigma}$ two coherent states) is given by \cite{phoenix} 
\begin{equation}
\hat{\mathcal{D}}_t^{j} |\lambda\rangle \langle \sigma| = \langle \sigma | \lambda\rangle^{\gamma \delta t} | \lambda e^{-\gamma \delta t} \rangle \langle \sigma e^{-\gamma \delta t}|.
\end{equation} 
In the limit $\delta t{\rightarrow}0$, $N{\rightarrow}\infty$ (so as to keep $t{=}N\delta{t}$ finite),  we get an accurate description of the dissipation-affected dynamics\cite{comment2}.  After the projection on the atomic part of the system, the state of the two mirrors is 
\begin{equation}
\begin{split}
\rho_m(t)&=\frac{1}{2}\big[\sum_{\mu=\alpha,\beta}|\mu_1(t),\mu_2(t) \rangle \langle \mu_1(t),\mu_2(t) |\\ 
&+ e^{-i \vartheta(t)-\Gamma(t) } |\beta_1(t),\beta_2(t) \rangle \langle \alpha_1(t),\alpha_2(t) |+h.c.\big] 
\end{split}
\end{equation}
where 
\begin{equation}
\begin{split}
&\alpha_j(t){=}\alpha_j e^{-\gamma t},~\beta_j(t){=}\alpha_j(t){+}(-1)^{j-1}i{\eta_j}(1{-}e^{-\gamma t})/{\gamma},\\
& \vartheta(t) = \sum_{j=1,2}({\eta_j}/{2\gamma})\alpha_j  (1-e^{-2\gamma  t}),\\
&\Gamma(t) =\sum_{j=1,2}({\eta_j^2}/2{\gamma^2})[\gamma t +\frac{1}{2}(1-e^{-2\gamma t}) - 2(1 - e^{-\gamma t})]. 
\end{split}
\end{equation}
The analysis of the CHSH inequality using $\rho_m(t)$ leads to features consistent with the solutions gathered through the Fokker-Planck approach. As the decoherence factor $\Gamma(t)$ grows with $(\eta/\gamma)^2$, the time-window where violation of the local realistic boundary can be observed gets smaller.

\section{Conclusions}

We have studied  a mediated coupling mechanism between a microscopic and a mesoscopic system in two different setup involving optical cavities with movable mirrors interacting with a three level atom.  The resulting dynamics drives the system  into  states which exhibit strong quantum features in both  cases considered. The study of  the first setup, involving a single mechanical oscillator and described in Sec. \ref{SM}, reveals strong non-local correlations between the atom and the movable mirror. Considerable violations of  Bell-CHSH inequality are observed  even when the thermal nature of the mirror's initial state is taken into account. Moreover, projective measurements over the atomic system probabilistically create non-classical mixed states of the mirror. Such non-classicality, quantified by the negativity of the Wigner function,  is robust against mechanical damping, while the dynamical mechanism we used ensures a good protection from other sources of noise. In the second part of the paper (Sec. \ref{TM}) a cavity where both mirrors oscillate around their equilibrium positions is considered. The conditional dynamics obtain by a post-selection process on the microscopic part of the system induces truly mesoscopic quantum correlations between the two mirrors which lead again to a violation of CHSH inequality at finite temperature. In analogy with the one-mirror setup, negative values of the Wigner function are found in the dissipative regime.

Apart from stimulating the experimental achievement of non-classical states of a massive system, which will be the focus of optomechanics at the quantum level, the first part of our proposal triggers the study of microscopic-mesoscopic interplay for mechanical manipulation and control. As a significant example, the bichromatic version of the coupling Hamiltonian opens up the interesting possibility to attach a non-trivial geometric phase to the state of the mechanical system. This can be done by adjusting the amplitude of displacements and the phase $\phi$ in a way so as to realize a cyclic evolution in the mirror's phase space, along the lines with Refs.~\cite{phase,Leibfried}. Such possibilities for microscopically-induced control of a mesoscopic device has already been studied elsewhere~\cite{noiGP} and it will be the topic of further investigations. The second part of our study focus on the quantum correlations shared by two massive objects, bringing our analysis to the boundary between the quantum and the classical world. In such operating conditions, the dissipative part of the dynamics induced by damping processes in the mechanical oscillators plays an important role. It is thus clear that the achievement of the condition $\eta\sim\gamma$ is crucial in our scheme, and a comment about the possibility of reaching this regime is unavoidable.  For state-of-the-art mechanical systems, typical values of $\gamma$ are in the range of a few Hz. On the other hand, the effective coupling rate $\eta$ is given by $\eta={\chi{g}^2\Omega^2}/{\delta^2\Delta^2},$ where $\chi$ is the radiation pressure interaction constant given by $\chi = (\omega_c /L) \sqrt{\hbar/2 m \omega_m}.$ For mechanical modes having  $\omega_m/(2\pi){=}300$KHz and mass $m \sim 50$ng placed to a cavity of $L = 10$mm ~\cite{Groblacher1,Groblacher2,Groeblacher} and assuming $g^2 \Omega^2 / \delta^2 \Delta^2 \sim 0.1$ and $\omega_c \sim 10^15$Hz,  a straightforward calculation shows that $\eta \sim 1.$ This  value is indeed  comparable to $\gamma$, thus demonstrating the achievability of the conditions required by our proposal. It is remarkable that the state of the two mechanical systems exhibits non-classical features both for one and two mirrors, in contrast with a purely optomechanical coupling between a movable mirror and a cavity field~\cite{mauro}. 

Our analysis demonstrates the broad validity of our arguments, both at the single and two-mirror level. We stress the full generality of our method. Although we have illustrated it using a specific setup, the same sort of quantum-correlated state can be engineered in settings consisting of a Bose-Einstein condensate in an optomechanical cavity,  two nano-mechanical resonators capacitively coupled to a Cooper-pair box or two planar superconducting resonators mutually connected via an off-resonant phase or transmon qubit~\cite{Noi2010,martinis,koch}. We hope that the results of our study to trigger experimental endeavors directed towards the achievement of the working conditions discussed here.

\section*{Acknowledgement}
We thank R. Fazio for valuable comments.  We acknowledge financial support from the National Research Foundation and Ministry of Education in Singapore, the UK EPSRC [through a Career Acceleration Fellowship (MP) and the ``New Directions for EPSRC Research Leaders" initiative, the Royal Society and the Wolfson Trust. VV is a fellow of Wolfson College, Oxford.

\appendix
\section{Adiabatic elimination}\label{appendix1}
We start from Eq. (\ref{HTot}) and we adiabatic eliminate the excited state of the atom $|e\rangle$ and the electromagnetic field inside the cavity. In order to do so, we assume  $\Delta >> \Omega, g$ and $\delta >> \Omega, g$.  We notice that the  only  terms in the Hamiltonian involving the atomic degrees of freedom are ${\cal H}_a$ and  ${\cal H}_R$. Hence, we perform first the adiabatic elimination of the exited level $|e\rangle$ of the atom. The Hamiltonian ${\cal H}_a + {\cal H}_R$ can be formally written as a $3\times3$ matrix with respect of the basis $\{|0\rangle,|1\rangle,|e\rangle\}$:
	 \begin{equation}
{\cal H}_a + {\cal H}_R= \begin{pmatrix} \label{atomH}
0 & 0& \Omega\\ 
0&0& g e^{i \Delta t} \hat{a}^\dag \\ 
\Omega & g e^{-i \Delta t} \hat{a} & \delta \end{pmatrix}.
	 \end{equation}
By writing a generic state of the atom as $|\lambda\rangle = c_0 |0\rangle + c_1 |1\rangle + c_e |e\rangle$ and by setting to zero $\dot{c}_e$ in the corresponding Schrodinger equation $i \partial_t |\lambda\rangle = ({\cal H}_a + {\cal H}_R) |\lambda\rangle$, we find the effective Hamiltonian 
	\begin{equation}
		\begin{split}
{\cal H}_{1}= &-\frac{\Omega^2}{\delta} |0\rangle \langle 0| - \frac{\Omega g e^{-i \Delta t} }{\delta} \hat{a} |0\rangle \langle 1|\\
&-  \frac{\Omega g e^{i \Delta t}}{\delta}  \hat{a}^\dag |1\rangle \langle 0| - \frac{g^2}{\delta} \hat{a}^\dag \hat{a} |1\rangle\langle 1|
		\end{split}
	\end{equation}
After the adiabatic elimination we substitute the terms ${\cal H}_a + {\cal H}_R$ in Eq. (\ref{HTot}) with the expression above and the total Hamiltonian of the system reads now ${\cal H}_{\rm sys} = {\cal H}_{eff}+ {\cal H}_c + {\cal H}_m +  {\cal H}_{mc} + {\cal H}_{cp}$

The next step is the elimination of the cavity field operators $\hat{a}$ and $\hat{a}^\dag$. In order to do so,  we consider the equations describing the time evolution of those operators $\dot{\hat{a}} = - i[H_{sys},\hat{a}]$ and $\dot{\hat{a}}^\dag = - i[H_{sys},\hat{a}^\dag]$ and we set to zero the time derivative. Considering that $[H_{sys},\hat{a}] = [H_c, \hat{a}] + [H_{mc}, \hat{a}] + [H_{eff}, \hat{a}]$, we find that
	\begin{equation}
		\begin{split}
[H_{sys}, \hat{a}] = &- \hat{a} \big(\Delta  + \chi_1 (\hat{b}^\dag_1 + \hat{b}_1) + \chi_2 (\hat{b}^\dag_2 + \hat{b}_2) - \frac{g^2}{\delta}  |1\rangle\langle 1|  \big)\\ 
& + \frac{\Omega g }{\delta} e^{i \Delta t} |1\rangle \langle 0|
		\end{split}
	\end{equation} 
Setting this quantity to zero and considering that $\Delta \gg \chi, g^2/\delta$, we find
	\begin{equation}
\hat{a} = \frac{\Omega g}{\delta \Delta} e^{i \Delta t} |1\rangle \langle 0|
	\end{equation}
In a similar way we find that 
	\begin{equation}
\hat{a}^\dag = \frac{\Omega g}{\delta \Delta} e^{-i \Delta t} |0\rangle \langle 1|
	\end{equation}
By substituting these equation in the expression for $H_{mc}$ we find the effective atom-mirrors interaction which reads as
	\begin{equation}
H_{am}^{eff} = \frac{\Omega^2 g^2}{\delta^2 \Delta^2}  \chi  |0\rangle \langle 0|(\hat{b}^\dag + \hat{b})
	\end{equation}
With $\eta = \frac{\Omega^2 g^2}{\delta^2 \Delta^2}  \chi$ we recover the expression in Eq. (\ref{HEffOneMirr}).


\begin{thebibliography} {99}
\bibitem{parameters} For $g\Omega/\delta\Delta\sim0.1$,  cavity-length $\sim0.3$mm and a light mirror (mass $\sim15$ng), $\chi\sim{2.8}$KHz, which corresponds to $\zeta\sim0.1$KHz. The elimination of $\ket{e}_a$  and the cavity field, whose damping rate is quenched by the off-resonant coupling, make the proposed dynamics quite feasible.
\bibitem{commentT0} Quite expectedly, the CHSH inequality is increasingly violated in time at $T=0$, asymptotically reaching Tsirelson's bound.
\end{thebibliography}
\end{document}